\documentclass[multphys,vecphys]{svmult}

\usepackage{makeidx}         
\usepackage{graphicx}        
\usepackage{multicol}        
\usepackage[bottom]{footmisc}

\makeindex             

\begin{document}

\title*{Good News for MOS, MXU \& Co. -- The New Spectroscopic Pipeline for the FORSes}
\titlerunning{The new spectroscopic FORS pipeline}
\author{Sabine Moehler
}
\institute{European Southern Observatory, Karl-Schwarzschild-Str. 2, 85748 Garching, Germany
\texttt{smoehler@eso.org}}
%
%
\maketitle
\begin{abstract}
Since October 1, 2006, spectroscopic data from the two FORS
instruments have been reduced with a new pipeline, which is based on a
bottom-up calibration approach. I give a short description of the pipeline and
discuss first experiences with automatic data reduction using this
software, which has significantly increased the percentage of processed
data for both instruments. I will also describe possible new options
for Quality Control.
\end{abstract}

\section{How does the Pipeline Work?}
\label{moehler:sec:pipe}
The spectroscopic pipeline for the FORS instruments can handle
long-slit and multi-object spectroscopic data using slitlets. In order
to be as flexible as possible most information is obtained directly
from the observed calibration data, minimizing the need for assumptions
(for details of the underlying principles see the contribution
by Izzo et al.).
\subsection{Calibration Data}
\begin{figure}
\centering
\includegraphics[height=0.99\textwidth,angle=270]{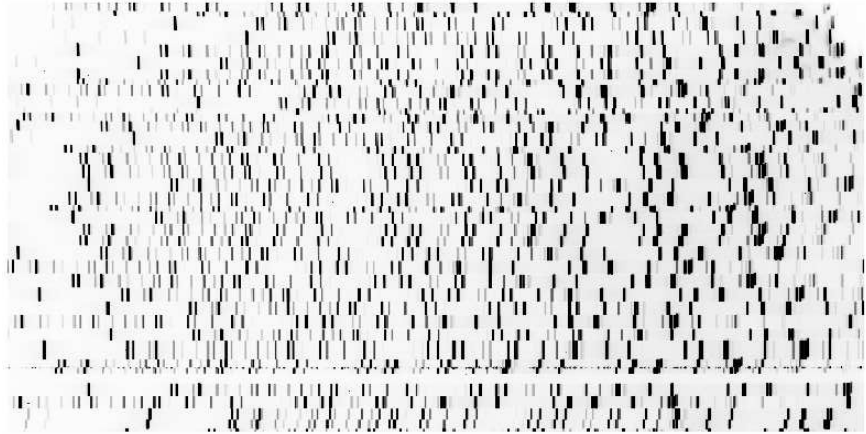}
\caption{Example of a raw arc frame for FORS2 MXU.}
\label{moehler:fig:arc_raw}       
\end{figure}

A first guess of the positions and lengths of the slitlets (or long
slit) is obtained from an arc lamp frame (see
Fig.~\ref{moehler:fig:arc_raw}). The positions of arc lamp spectra on
the CCD are determined by a pattern matching technique applied to the
detected emission lines. For that an estimate of the linear dispersion
as well as a line catalog (which may both be provided by the user) are
required. Adjacent slitlets with no offsets in dispersion direction
are not distinguished and will form one longer slitlet.

\begin{figure}
\centering
\includegraphics[width=0.5\textwidth,angle=0]{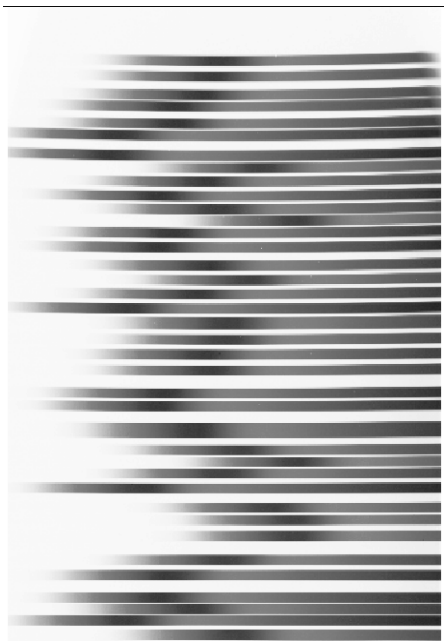}
\caption{Example of spatial curvature for a raw FORS2 MXU flat field.}
\label{moehler:fig:spat_curv}       
\end{figure}

The spatial curvature is determined by tracing the edges of the
corresponding flat field spectra
(cf. Fig.~\ref{moehler:fig:spat_curv}) when possible. The curvature is
fit by default with a second order polynomial.  The final wavelength
calibration is obtained taking into account the spatial curvature of
the spectra.

The pipeline provides the following output calibration files:
\begin{itemize}
\item slit positions on the CCD at the central wavelength of the grism
\item coefficients for the dispersion relation and the spatial curvature
\item master bias
\item master flat field (normalized and not-normalized)
\item wavelength map and spatial map
\item reduced arc lamp frame
\item spectral resolution and line widths
\end{itemize}

The wavelength map specifies at the position of each original CCD
pixel its corresponding wavelength. Thus the user can avoid the
re-sampling of the spectra usually done after the wavelength
calibration. The spatial map provides in the same manner the position
of the CCD pixel within its associated slitlet. The reduced arc lamp
frame allows to judge the quality of the calibration (straight arc
lines, no wriggles, no offsets). If successful (see
Sect.~\ref{moehler:sec:weak}) the pipeline achieves an accuracy of 0.1
pixel both in spatial and dispersion direction.

\subsection{Science Data}
The default reduction of spectroscopic science data is as follows: The
data are first corrected for bias. Then the extraction mask derived
above (i.e. the positions of the slitlets and the curvatures of the
resulting spectra) is applied to the science data; they are
flat-fielded and re-mapped eliminating the optical
distortions. Afterward they are rebinned to constant wavelength
steps. The wavelength calibration can be adjusted using sky emission
lines. This allows to correct for shifts between night-time science
and day-time calibration data. Such shifts (of the order of 1 pixel)
are known to happen due to flexure for MXU observations. Finally
object spectra are detected and extracted, together with the corresponding
error spectra and sky spectra. For long-slit spectroscopy the sky
background is determined using a median. This is a valid approach if
not more than half of the pixels contain flux from objects {\em and}
there is no spatial gradient in the sky background. For multi-object
spectroscopy instead the sky is subtracted before remapping, i.e. when
the spectra are still in the original CCD coordinate system. The sky
is determined with a robust linear fitting, which allows for a linear
spatial gradient in the background. Also in this case, however, not
more than half of the pixels may contain flux from the object(s).

The output science files are
\begin{itemize}
\item positions and widths of object spectra
\item extracted object spectra, error spectra, and sky spectra
\item unmapped  (corrected for spatial distortions,
  not rebinned) science frames 
\item mapped (i.e. corrected for spatial distortions
  and rebinned) science frames 
\item adjusted dispersion coefficients and wavelength map
\end{itemize}
\section{Strengths and Weaknesses}\label{moehler:sec:weak}
The major strength of the new pipeline is its flexibility and
robustness. It requires only very limited input information (mainly
dispersion estimate and line catalog) and can therefore be applied to
a large variety of instrument configurations. Thus it is now possible
for the first time to {\em automatically} reduce long-slit (LSS) and
multi-object spectroscopic data (MOS/MXU) from the two FORS
instruments for all grisms and both collimators (standard and
high-resolution). To verify its flexibility the pipeline was applied
to data from the Low-Resolution Spectrograph of the Hobby Eberly
Telescope, which it handled without any problems.

\begin{figure}
\centering
\includegraphics[height=0.45\textwidth,angle=0]{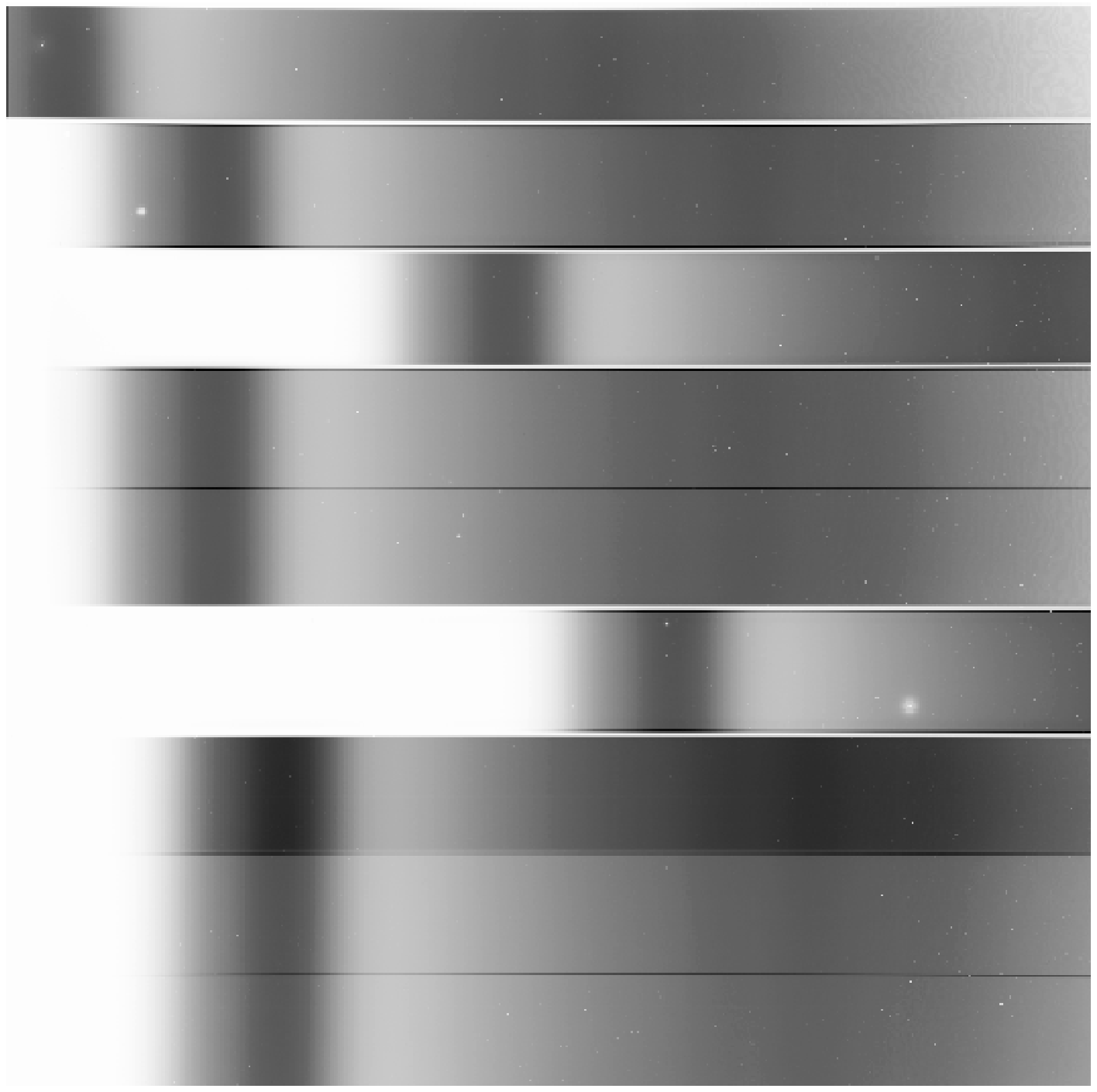}
\includegraphics[height=0.45\textwidth,angle=0]{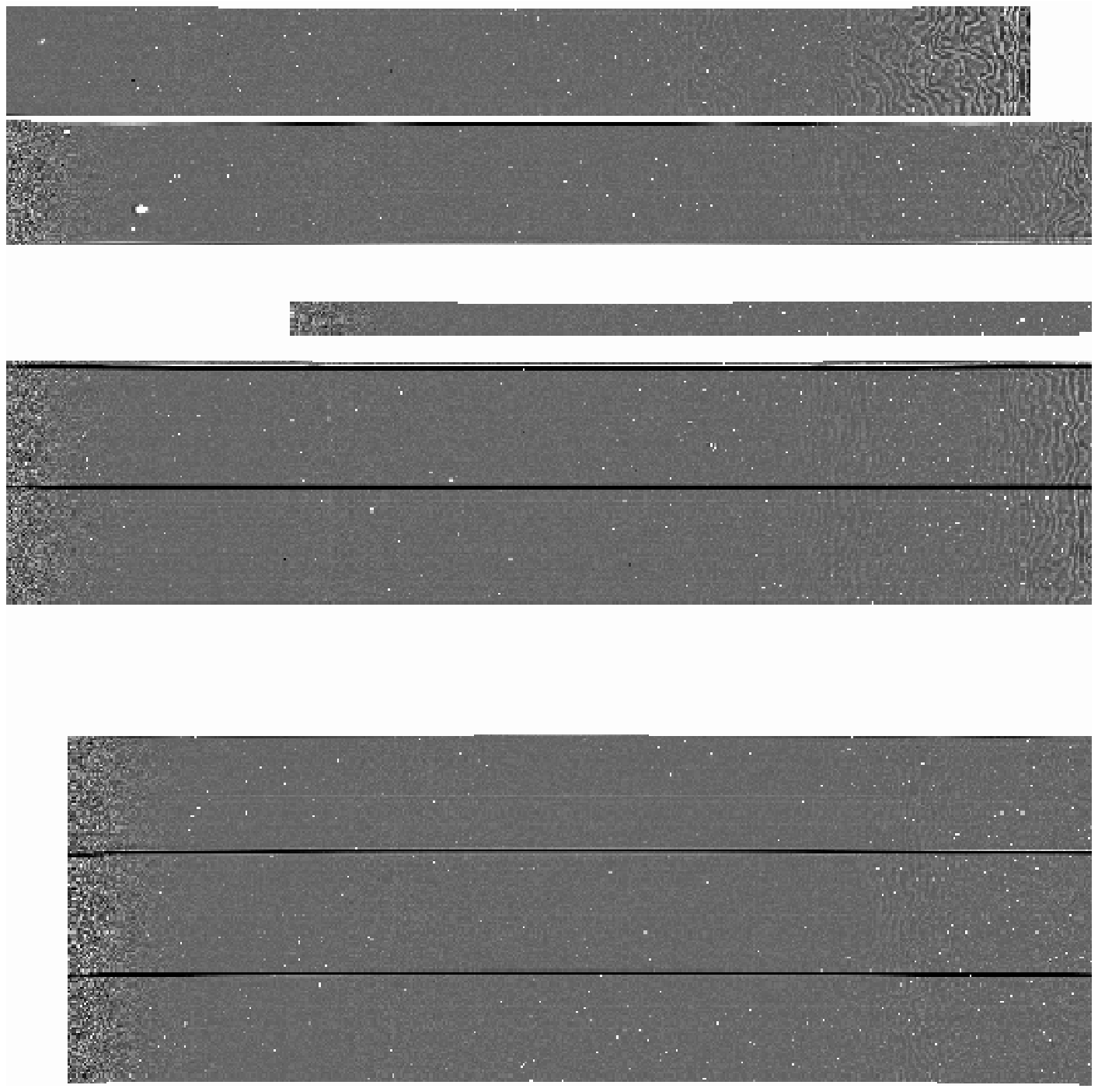}
\caption{Example of missed slitlets in FORS2 MOS data (left: averaged
screen flat, right: normalized screen flat).}
\label{moehler:fig:bad_mos}       
\end{figure}

The pipeline does have problems if there are too few arc lines,
e.g. for slitlets with large offsets. In such cases it sometimes fails
to correctly identify the slitlet. This can be most easily recognized
by comparing the not-normalized and the normalized screen flat
(cf. Fig.~\ref{moehler:fig:bad_mos} -- one slitlet is partly missing,
another one was not processed at all). Large gaps between arc lines
can be problematic as non-linear dispersion terms can become
important, while the pipeline currently uses only linear dispersion
estimates for the pattern matching.

The pipeline will fail in case of regularly spaced arc lines, like
those created by a laser comb, as its pattern matching (see
contribution by Izzo et al.) does not work for such data.

\section{Quality Control}
In order to check the quality of observational data a pipeline is of
great importance. Otherwise it is difficult to distinguish between
well-known and correctable instrument effects and real
problems. Quality control shall ensure that the data observed with the
FORSes can be calibrated. In addition the trending of certain
parameters allows to have an eye on the instrument health (see
http://www.eso.org/dfo/quality for more details). Therefore the
resolution, central wavelength, and number of identified arc lines are
regularly monitored. A change in the number of identified lines can
indicate reduced flux of the arc lamp and possibly impending failure
of the lamp. A change in resolution could indicate focus problems.

\begin{figure}
\centering
\includegraphics[height=\textwidth,angle=270]{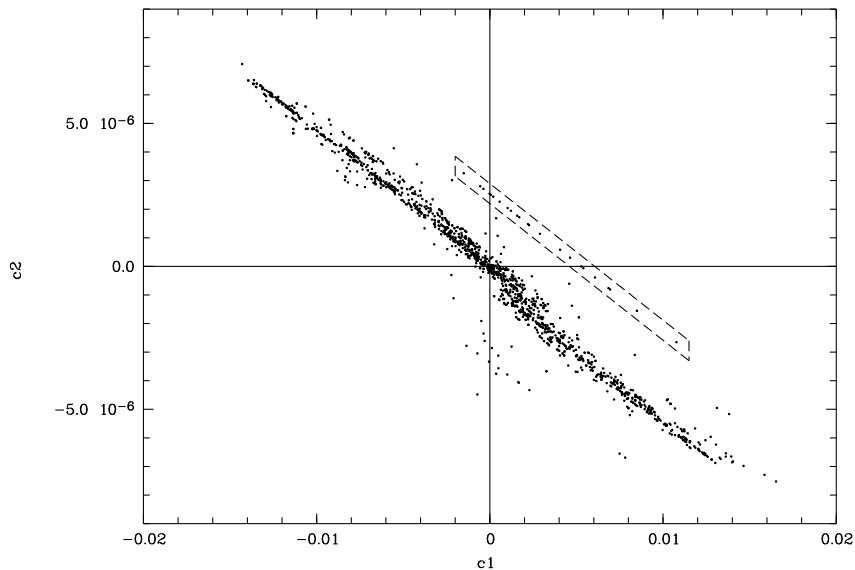}
\caption{Curvature coefficients for FORS2 MXU data for the 600z
grism. The lower branch resulting from chip 2 data indicates that chip
2 is mounted at an angle of 0$\stackrel{\circ}{.}$079$\pm$
0$\stackrel{\circ}{.}$007 with respect to chip 1 (the known value being
0$\stackrel{\circ}{.}$083). The reduced arc
lamp image corresponding to the coefficients in the dashed box is
shown in Fig.~\ref{moehler:fig:PCCX_bad}}
\label{moehler:fig:coeff}       
\end{figure}
The situation for the quality control of FORS data has vastly improved
with the new pipeline, as now almost all data can be reduced (with
polarimetric data being the only exception).  Moreover, the new
pipeline also provides additional information on the instrument. For
example, the combination of the curvature coefficients can be used to
monitor grism alignment: The coefficients c1 and c2 describe the slope
and the curvature of a slitlet, respectively. If the curvature is 0,
then the slope should be 0 as well -- otherwise this indicates that
the grism is not well aligned. Fig.~\ref{moehler:fig:coeff} shows the
coefficients obtained for the 600z FORS2 grism. Obviously the
correlation for the majority of solutions passes through zero.

\begin{figure}
\centering
\includegraphics[width=\textwidth,angle=0]{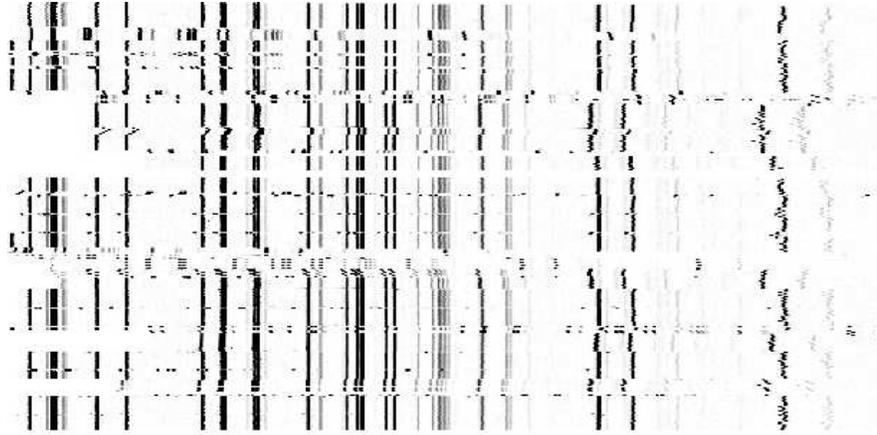}
\caption{Reduced arc lamp frame for the deviating points marked in
Fig.~\ref{moehler:fig:coeff}.}
\label{moehler:fig:PCCX_bad}       
\end{figure}

Some solutions, however, show significant offsets from the general
trend. Fig.~\ref{moehler:fig:PCCX_bad} shows a reduced arc lamp image
for such a deviating case (marked by the dashed box in
Fig.~\ref{moehler:fig:coeff}). Obviously the wavelength calibration
obtained for several slitlets is rather bad (wriggles instead of
straight lines). Thus diagrams like Fig.~\ref{moehler:fig:coeff} allow
to look for bad solutions efficiently in case of large data volume,
which may prohibit checking all solutions individually.

Other new parameters become available if there are more than 12
slitlets distributed across the CCD. In this case the pipeline
calculates a global distortion model, which contains good indicators
for the monitoring of the instrument's health, like the
instrument scale.

\section{Concluding Remarks}
The new spectroscopic pipeline for the FORS instruments has been a
great success in that it allows to reduce {\em automatically}
long-slit and multi-object spectroscopic data for all FORS grisms. The
fact that it does not need a first guess from an instrument model
makes it a promising candidate for the reduction of spectroscopic data
from a variety of instruments. In addition this approach is extremely
helpful in the presence of instrument instabilities as it allows to
reduce data whose configurations deviate from the expected one
(e.g. due to instrument aging or earthquakes).

\section*{Acknowledgments}
I would like to thank C. Izzo and U. Hopp for their valuable comments on this
manuscript.

\printindex
\end{document}